\documentclass[aps,prl,showpacs,twocolumn,groupedaddress]{revtex4-1}
\usepackage{graphicx,graphics,color,epsfig}
\usepackage{amsmath}
\usepackage{amssymb}
\usepackage{amsfonts}
\usepackage{amsfonts}
\usepackage{epstopdf}
\usepackage{bm}
\usepackage{times,xspace}
\usepackage{color}

\begin{document}

%\preprint{PRL}

\title{Electron-like Fermi surface and in-plane anisotropy due to chain states in YBa$_2$Cu$_3$O$_{7-\delta}$ superconductors}

\author{Tanmoy Das}%\affil{1}{Theoretical Division, Los Alamos National Laboratory, Los Alamos, NM 87545, USA}}
\affiliation{Theoretical Division, Los Alamos National Laboratory, Los Alamos, New Mexico 87545 USA.}

\date{\today}

\begin{abstract}
%We present a novel explanation for the emergence of electron-like Fermi surface in the pseudogap phase in  YBa$_2$Cu$_3$O$_{7-\delta}$ (YBCO) materials as demonstrated by numerous magneto-transport measurements.
We present magneto-transport calculations for YBa$_2$Cu$_3$O$_{7-\delta}$ (YBCO) materials to show that the electron-like metallic chain state gives both the negative Hall effect and in-plane anisotropic large Nernst signal. We show that the inevitable presence of the metallic 1D CuO chain layer lying between the CuO$_2$ bilayers in YBCO renders an electron-like Fermi surface in the doping range as wide as $p=0.05$ to overdoping. With underdoping, a pseudogap opening in the CuO$_2$ state reduces its hole-carrier contribution, and therefore the net electron-like quasiparticles dominate the transport properties, and a negative Hall resistance commences. We also show that the observation of in-plane anisotropy in the Nernst signal -- which was taken as a definite evidence of the electronic `nematic' pseudogap phase -- is naturally explained by including the `quasi-uniaxial' metallic chain state. Finally, we comment on how the chain state can also lead to electron-like quantum oscillations.
\end{abstract}


\maketitle

An understanding of the pairing mechanism in cuprate superconductors relies on the underlying complex Fermi surface (FS) properties, and how they are derived from the normal state pseudogap (PG) phase. Important clues to the origin of the PG state have been put forward by recent Hall effect measurements which observed a negative Hall resistance below the PG temperature in YBCO systems.\cite{HallEP} This result received further supports from the negative Seebeck coefficient,\cite{seebeck,seebeckNC} strongly enhanced Nernst effect,\cite{Nernst_nematic,Nernst_data} as well as identification of quantum oscillations\cite{SdH2007,SdH2008,sebastian_compensated,sebastianmass,QOPDai} with cyclotron frequency indicative of electron-like FS. These stimulating discoveries, which according to the conventional Onsager-Lifshitz paradigm\cite{Onsagar} implies a closed electron pocket, point toward the topological changes in the FS in the PG state when compared with its large metallic FS in the overdoped sample.\cite{ARPES_damascelli} Such small electron pockets do not arise naturally from the band-structure calculations considering the CuO$_2$ planes and their total area is inconsistent with the nominal doping concentration.\cite{LDA}

To explain these results, candidate proposals for various symmetry-breaking patterns have been offered, leading to drastic FS reconstruction into multiple Fermi pockets of different natures.\cite{Norman,Norman_Lifshitz,DDW,DasSDW,fldCO,Harrison,Sachdev_SDW,DDW_Nernst} We would, however, expect to see signatures of these pockets in spectroscopies such as angle-resolved photoemission spectroscopy (ARPES) and scanning tunneling microscopy/spectroscopy (STM/S). Yet, extensive ARPES\cite{ARPES_damascelli,ARPES_borisenko}, and STM\cite{STM_FA} studies on various cuprate materials consistently have so far been able to map out only one Fermi pocket or `Fermi arc' -- both would represent the same FS segment if the intensity on the pocket's back side is suppressed by the coherence factors -- which is again of hole-like character.\cite{DDW,DasSDW} Furthermore, to explain the large negative Hall coefficient, one requires to have electron-like quasiparticles dominating over the hole quasiparticles on the FS, which is prohibited in hole doped system due to Luttinger theorem. On the basis of this reasoning, the recent experimental evidence of a field induced charge ordering\cite{fldCO} will also find difficulty in explaining the negative Hall coefficient even if it is assumed to induce electron pocket.\cite{Harrison} Taken together, the electron-pocket scenarios of various density wave origins\cite{Norman,DDW,DasSDW,Harrison} are inadequate to explain the negative slope of the magneto-transport data.

In this study, we introduce a fundamentally new perspective -- we show that the emergence of the electron-like FS in the PG state can be rationalized by realistically considering the contributions of the much ignored metallic CuO chain bands in YBCO. The unambiguous existence of the chain states on the FS has been well established by many ARPES studies from extreme underdoped region ($p=0.05$) to overdoping ($p=0.29$ and above).\cite{ARPES_damascelli,ARPES_borisenko} Its contribution to $c$-axis transport\cite{Hussey} and quantum oscillation in specific heat\cite{Riggs} measurements is also studied earlier, further supporting the metallic character of chain state. The chain states seem to exhibit neither the PG opening nor any considerable doping dependence. Nevertheless, it dominates in the electronic spectra in the PG state due to its interplay with the CuO$_2$ states. Our reasoning that underlies this conclusion is that the PG opens in the CuO$_2$ bilayers, truncating its full metallic FS into small segments. Irrespective of the specific origin and nature of the small FS, it contains much reduced hole-carrier concentrations, determining the nominal hole doping of the sample. Therefore, the low-energy electronic state becomes effectively characterized by electron-like charge carriers of the chain state. Above the PG regime both in temperature ($T$) and doping ($p$), the large CuO$_2$ state is recovered, restoring the predominant hole-carrier on the FS of YBCO. The resulting two-carriers FSs quantify the negative Hall effect,\cite{HallEP} and also the enhanced Nernst signal with prominent in-plane anisotropy\cite{Nernst_nematic,Nernst_data} due to the `quasi-uniaxial' character of the chain state.

{\it Tri-layer lattice model:-} Encouraged by the ARPES results\cite{ARPES_damascelli,ARPES_borisenko} and the first-principle calculations,\cite{LDA} we formulate a unified non-interacting three band model, originating from 2D CuO$_2$ bilayers and an 1D CuO chain.\cite{atkinson,footortho} To effectively model the reminiscent nodal FS segment of the CuO$_2$ bands in the PG state, we consider some form of density-wave order whose front side would also accommodate the `Fermi arc' property. Constrained by this assumption, a Hartree-Fock like order parameter is introduced to the Hamiltonian with a periodic modulation ${\bm Q}=(\pi,\pi)$, which is commensurate with the CuO$_2$ lattice plane. We can think of a residual spin-\cite{DasSDW} or charge-density wave\cite{fldCO} (S/CDW) or a $d$-density wave\cite{DDW} or some other commensurate FS instability as the microscopic origin of such ordering, however, the macroscopic properties we aim to describe here do not rely on these details.\cite{DasSDW} On the basis of these experimental considerations, we deduce the general form of the Hamiltonian in the Nambu representation per unit cell:

\begin{figure}%[top]
%\hspace{2cm}
\centering
\rotatebox{0}{\scalebox{.35}{\includegraphics{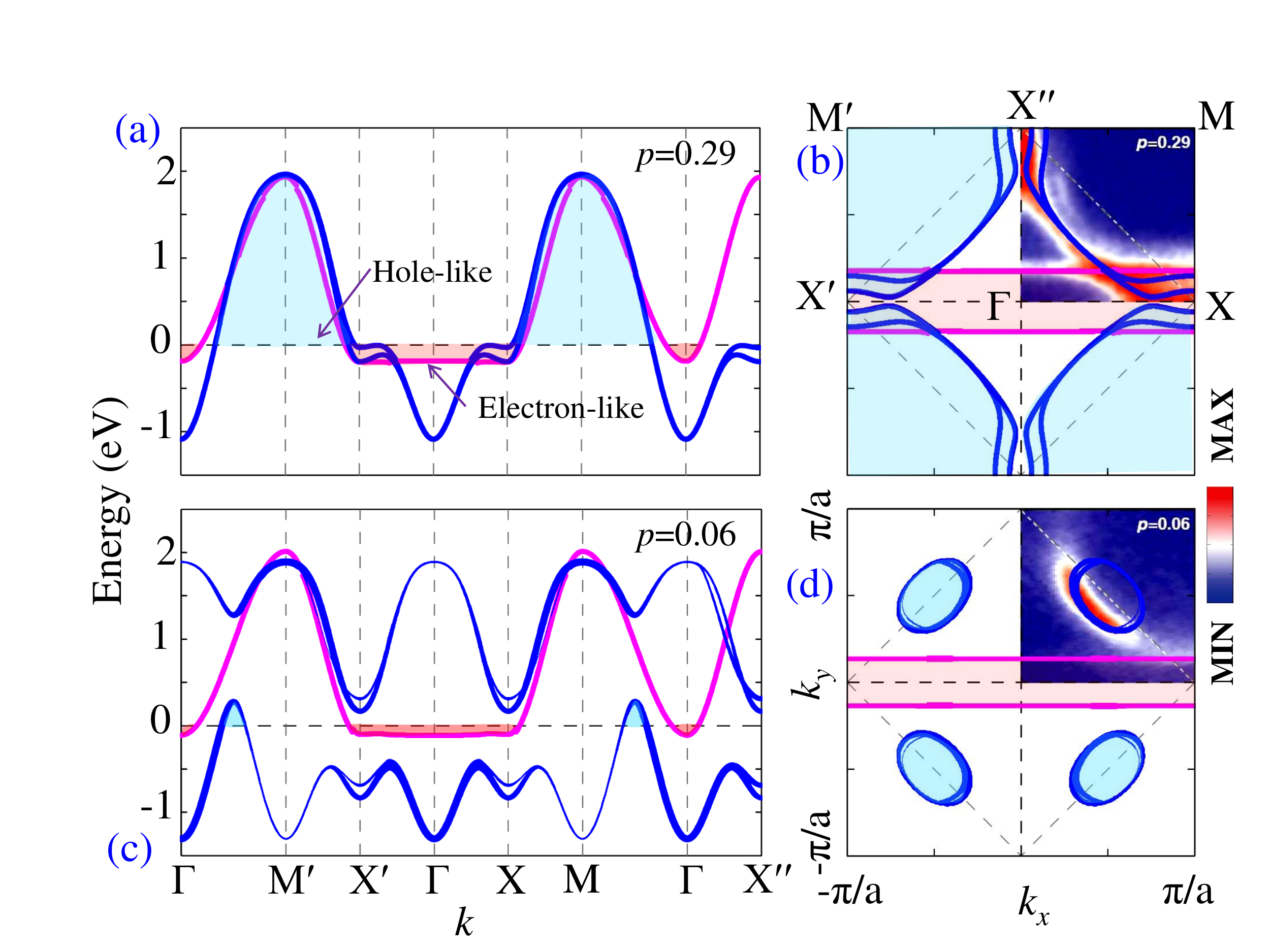}}}
\caption{(Color online) %{Electronic dispersion and Fermi surface of tri-layer lattice model.}
(a) Paramagnetic dispersion obtained from the eigenstates of the Hamiltonian in Eq.~\ref{Ham}. Blue lines are the bonding and antibonding bands arising from the CuO$_2$ planes, while magenta line depicts the chain state. The light blue and red shadings represent the hole-like and electron-like character of the bands, respectively. (b) Corresponding computed FSs are compared with the ARPES data\cite{ARPES_damascelli} (plotted only in one quadrant of the Brillouin zone for visualization).  (c) Same as in (a), but computed in the PG state in an underdoped region. The width of each line gives the associated spectral-weight, determined by the SDW coherence factors. (d) FSs in the PG state host only tiny hole pockets centering nodal points.}
\label{fig1}
\end{figure}

\begin{eqnarray}
H = \sum_{{\bm k}\sigma,ij}\left[\xi_{ij{\bm k}}c_{i{\bm k}\sigma}^{\dag}c_{j{\bm k}\sigma}+ V_{ij{\bm k}}c_{i({\bm k}+{\bm Q})\sigma}^{\dag}c_{j{\bm k}\bar{\sigma}} + h.c. \right],
\label{Ham}
\end{eqnarray}
where $c_{i{\bm k}\sigma}^{\dag}~(c_{i{\bm k}\sigma})$ creates (annihilates) an electron with momentum ${\bm k}$ and spin $\sigma$ on the $i=p,p^{\prime},c$ tri-layers. The singe-particle dispersions $\xi_{ij}$ are defined within tight-binding expansion of the Cu $d_{x^2-y^2}$-orbitals (hybridized intrinsically with O $p$ orbitals) hopping on a periodic lattice in the tetragonal crystal as given in Ref~\onlinecite{footTB}.
%
%\begin{eqnarray}
%%
%&&\xi_{pp} =-2t(\phi_x+\phi_y) -4t^{\prime}\phi_x \phi_y -2t^{\prime\prime}(\phi_{2x}+\phi_{2y})-\mu_p,\nonumber\\
%%
%&&\xi_{cc} =-2t_{c}\phi_y-\mu_c, \nonumber\\
%%
%&&\xi_{pp^{\prime}} =-2t_{pp^{\prime}}(\phi_x-\phi_y)\phi_{z/4},~~~\xi_{cp} =-2t_{cp}\phi_{z/4},
%%
%\label{disp}
%\end{eqnarray}
%
%where $\phi_{\alpha \nu}$=$\cos{(\alpha k_{\nu})}$ with $\nu$=$x$,$y$,$z$. $\mu_{p,c}$ are the chemical potential for the plane and chain states, respectively, which encode relative onsite energy differences and other crystal effects between the two levels. $t$,$t^{\prime}$, and $t^{\prime\prime}$ are the 1st, 2nd and 3rd nearest neighbor (NN) hopping matrix-elements on the 2D CuO$_2$ plane, $t_c$ is the NN hopping on the 1D chain aligned along $y$-axis, and $t_{pp^{\prime}}~(t_{cp})$ is the plane-plane (chain-plane) tunneling matrix-element along $c$-axis. We benchmark the values of the hopping integrals by fitting the eigenstates of the non-interacting part of the Hamiltonian (setting $V$=0) to the experimental FSs (see Fig.~\ref{fig1}{a}) and to the first-principle deduced dispersions of Ref.~\cite{LDA}. The chemical potentials are evaluated self-consistently to maintain the doping concentration, $p$, of the CuO$_2$ bands.

Next, we exemplify how a small hole-pocket forms in the PG state. To simplify the analysis, while retaining the salient spectroscopic features, we adopt a ${\bm k}$-independent interaction potential $V$ to be same for both bilayers states, while setting $V=0$ for the chain layer, as the latter do not exhibit any PG opening. We set the spin index $\bar{\sigma}=-\sigma$ to construct a SDW modulation. %The evidence of SDW order has been recently obtained in the PG region via spin-polarized neutron diffraction in YBCO\cite{MookN} and Hg-based cuprates\cite{GrevenN} and muon spin-relaxation measurements.\cite{MookmuSR} 
In this case, $V$ encodes the onsite Hubbard $U$, and a order parameter $S$, as $V_i=US_i$, with $S_i=\langle c^{\dag}_{i({\bm k}+{\bm Q})\sigma}c_{i{\bm k}\sigma}\rangle$ ($i=p,p^{\prime}$), where the canonical average is taken over the entire Brillouin zone.  We use the value of effective $U$=1.59 eV, deduced from earlier calculations,\cite{DasSDW,Dahm} and the order parameter $S$ is evaluated self-consistent at each $T$ and doping. See supplementary material\cite{SM} for details. We emphasize that our choice of a particular form of density wave does not preclude us from analyzing the full problem, rather improves the reliability of the results with respect to a single parameter $U$. The choice of ${\bm Q}=(\pi,\pi)$ order however gives the FS topology which agrees well with most ARPES data.

\begin{figure}%[top]
\hspace{-0cm}
\centering
\rotatebox{0}{\scalebox{.32}{\includegraphics{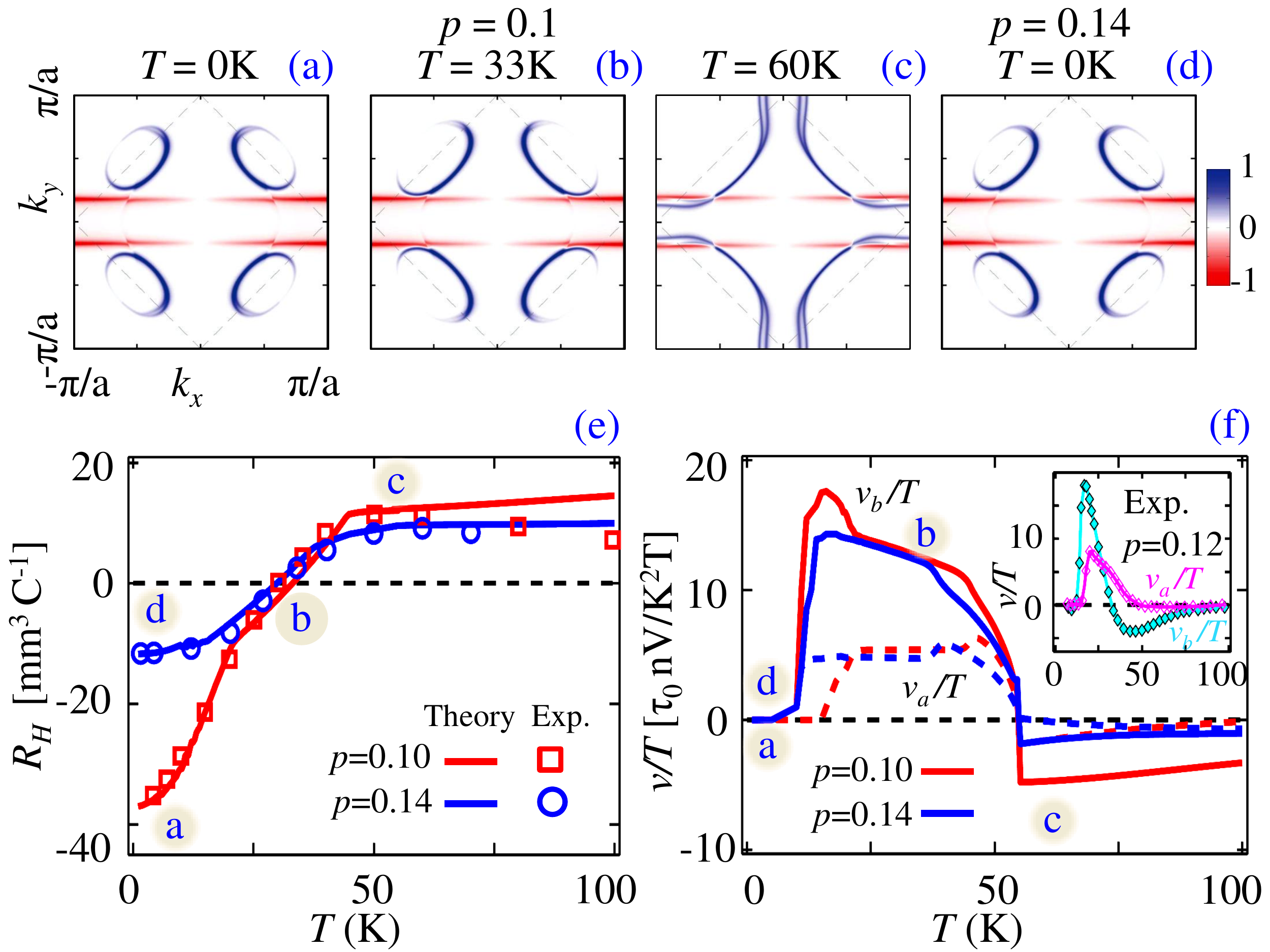}}}
\caption{(Color online) %{`Quasi-charge-density' and Hall and Nernst coefficients.}
(a)-(d) 2D color plots of the total `quasi-charge-weight' $\mathcal{C}_{ij}({\bm k},E_F)$ (averaged over all layers) plotted at four representative cases. The red-white-blue color scheme gives the electron-like quasiparticle to no quasiparticle to hole-like quasiparticle states at a given ${\bm k}$ on the FS. (a) `Quasi-charge-weight' calculated at an extreme low-$T$ region for doping $p=0.10$, where the Hall coefficient, $R_H$ is negative and Nernst coefficient, $\nu/T$, is small, demonstrating the dominance of the electron-like quasiparticle state on the FS. (b) is chosen at an intermediate $T$ where `quasi-electron' and `quasi-hole' weight compensate each other to lead to zero Hall coefficient and enhanced Nernst signal. At a higher $T$ in (c) SDW disappears and the `quasi-hole' weight dominates. Here Hall  and Nernst coefficients flip sign. The plot in (d) is similar to that in (a), but at a higher doping $p=0.14$. It should be noted here that, unlike in a strict ambipolar state, the maximum of the Nernst signal shifts from the zero of Hall coefficient, due to overlapping FS topology and matrix-element effects. (e) The computed $T-$evolution of the Hall coefficient, $R_H$, is compared with corresponding experimental data\cite{HallEP} at two representative dopings of YBCO. The excellent agreement below the SDW temperature can be clearly marked. (f) Corresponding computed $T$-evolution of the Nernst coefficient, $\nu_a=\nu_{xy}$ and $\nu_b=\nu_{yx}$, at same $T$ as in (e).  The experimental data for YBCO (shown in inset) is available at a slightly different doping.\cite{Nernst_data} Despite a characteristically good agreement between theory and experiments, we expect discrepancies due to the relaxation time approximation as mentioned in the text. Results of $\nu_a$ and $\nu_b$ reveal a large anisotropy, which is $T$-dependent. The source is this anisotropy is the presence of the `quasi-uniaxial' chain state, lying along the $y$-axis of the crystal. The anisotropy is strongest when the Nernst signal attains its maximum, in good agreement with experimental data taken from Ref.~\cite{Nernst_data}}
\label{fig2}
\end{figure}

In Figs.~\ref{fig1}(c) and (d), we notice that a direct band gap is opened at $E_F$ everywhere in the Brillouin zone, except at the nodal points, giving rise to a hole-pocket in the CuO$_2$ states. The width of each line depicts the associated quasiparticle weight which is in detailed agreement with experiments, given that the weak intensity of the shadow bands can be detected by ARPES.\cite{JMesot,ZXshadow} It is important to note that while the parameters of the chain states are doping independent, its electronic states posses a weak doping evolution via the interlayer coupling to the plane states. It is self-evident from the band curvature of the chain states at $E_F$ that it continues to host electron-like FS (although open orbit) at all doping; see Fig.~\ref{fig2} for further illustration. The electron-like property of the chain state is also evident in the ARPES data of Ref.~\onlinecite{ARPES_damascelli}.

{\it Hall and Nernst effect calculation:-} Relying on the aforementioned effective normal state Hamiltonian for YBCO, we now proceed to study the quasi-particle transport properties based on a semiclassical numerical calculation. Nernst experiment measures the transverse electric field, ${\bm E}$, in response to a combination set-up of an externally-imposed temperature gradient, $\nabla T$, and an orthogonal magnetic field, ${\bm B}$.\cite{Nernst_nematic,Nernst_data}
% which vanishes the net electrical current flow ${\bm J}$. []
The net electric current density ${\bm J}$ produced via this effect is essentially related to ${\bm E}$ and $\nabla T$ via linear response in-plane Hall conductivity tensor $\hat{\sigma}$ and Nernst conductivity tensor $\hat{\alpha}$, as ${\bm J}$=$\hat{\sigma}{\bm E}$-$\hat{\alpha}\nabla T$. In case of a perfect cancelation of the net charge current, we obtain a working formula for these two measured quantities:
%
%\begin{equation}
%
${\bm E}=\hat{\sigma}^{-1}\hat{\alpha}\nabla T=-\hat{\theta}\nabla T.$
%
%\end{equation}
%
Experimentally, the Hall resistance, $\hat{\rho}$=$\hat{\sigma}^{-1}$, and the Nernst coefficient, ${\nu}$=$\hat{\theta}/B$, are measured independently via transport probes typically by setting a weak magnetic field ${\bm B}$=$B\hat{z}$ along the $c$-axis of the lattice.\cite{HallEP,Nernst_nematic}

In theory, $\hat{\sigma}$ and $\hat{\alpha}$ are calculated by solving the standard Boltzmann equations for the low-$T$ DC transport.\cite{Sachdev_SDW,DDW_Nernst} After some tedious algebra we derive the formalism for these quantities in terms of the quasiparticle states and their associated weight (see supplementary material\cite{SM}):
\begin{eqnarray}
\hat{\sigma}_{\mu\nu}^{ij}=\frac{\beta e^2}{2}\sum_{{\bm k},n}M_{n{\bm k}}^{ij}v^{\mu}_{n{\bm k}}\Omega_{n{\bm k}}^{-1}v^{\nu}_{n{\bm k}}{\rm sech}^2\left(\frac{\beta E_{n{\bm k}}}{2}\right),\hspace{22pt}\\%\nonumber\\
\label{sigma}
\hat{\alpha}_{\mu\nu}^{ij}=-\frac{\beta^2 e}{2}\sum_{{\bm k},n}M_{n{\bm k}}^{ij}v^{\mu}_{n{\bm k}}\Omega_{n{\bm k}}^{-1}v^{\nu}_{n{\bm k}}E_{n{\bm k}}{\rm sech}^2\left(\frac{\beta E_{n{\bm k}}}{2}\right).%\nonumber\\
\label{alpha}
\end{eqnarray}
Here $E_{n{\bm k}}$ stands for the quasiparticle band with wavevector ${\bm k}$, and band index $n$, of Hamiltonian in Eq.~\ref{Ham}. Denoting the corresponding Bloch eigenstate by $\psi^{i}_{n{\bm k}}$, we express the transport matrix-element $M_{n{\bm k}}^{ij}=\psi^{i}_{n{\bm k}}\psi^{j\dag}_{n{\bm k}}$. $M$ essentially projects the transport tensors from the Bloch quasiparticle representation to the tri-layer lattice indices $i,j$. $v^{\mu}_{n{\bm k}}=\partial E_{n{\bm k}}/\hbar\partial{\bm k}_{\mu}$ is the quasiparticle velocity along the $\mu=x,y,z$ axes, and $\beta=1/k_BT$, where $k_B$ is Boltzmann constant. The differential operator that drives the system into a non-equilibrium state in response to the applied field is evaluated within relaxation-time approximation\cite{RelaxA}: $\Omega_{n{\bm k}}=\left[-\frac{e}{\hbar c}({\bm v}_{n{\bm k}}\times {\bm B})\cdot\nabla_{\bm k}+\tau^{-1}_{\bm k}\right]$, where the relaxation time is taken to be constant both in ${\bm k}$ and $T$.\cite{foottau} $e$, $\hbar$ and $c$ have their usual meanings.

The accurate evaluation of the $T$-dependence of the transport properties requires an experimentally constrained $T$-dependent gap parameters, in other words, an accurate description of the $T$-evolution of the FS topology. It is well known that a mean-field order parameter overestimates the PG transition temperature.\cite{DasSDW,Sachdev_SDW} To overcome this difficulty, we adopt a phenomenological form of the $T$-dependence of the gap parameter $V(T)$=$V\sqrt{1-T/T_{s}}$,\cite{Sachdev_SDW,TdepPG} with the onset $T$ for the density wave order is set to be $T_{s}\approx55$~K, in close agreement with Neutron measurement for Nd-doped La-based cuprate\cite{Ts}. However, at each $T$, the Fermi level is evaluated self-consistently to preserve the hole counting.

{\it Results:-} Fig.~\ref{fig2} displays the $T-$evolution of the Hall coefficient $R_H=\rho_{yx}/B$ and Nernst coefficient $\nu_{yx}/T$ at two dopings at which the normal state Hall coefficient data are available (the value of $B$ is scaled out in the presented results). A visual comparison between theory and experiment reveals a systematic agreement between them. To relate the sign and magnitude of $R_H$ and $\nu_{yx}/T$ to their corresponding charge carrier features, we introduce a function called `quasi-charge weight', in analogy with the quasi-particle weight: $\mathcal{C}_{ij}({\bm k},\omega=E_F)=\sum_{n,{\bm k}}e_{n{\bm k}}\psi^{i}_{n{\bm k}}\psi^{j\dag}_{n{\bm k}}\delta(\omega-E_{n{\bm k}})$, where $e_{n{\bm k}}=\pm e$ at electron and hole like $k_F$ obtained from the slope of the bands at $E_F$, respectively, for $n^{th}$ band. We emphasize that in the orthorhombic real-space unit cell, the Hall current arises due to finite tunneling between two adjacent chain layers (which gives a warped chain FS in the $k$-space). We plot the total the `quasi-charge weight' in Fig.~\ref{fig2}{a} to {d} for several representative cases where the Hall coefficient is negative, zero and positive as a function of $T$ and doping.

The obtained results confirm our aforementioned postulates that at low-$T$ when the strong PG shrinks the CuO$_2$ states into small hole-pockets, the dominant chain states produce a net electron-like quasiparticle character on the FS. With increasing $T$ when the strength of the PG reduces, the area of the hole-pocket gradually increases, while the chain state remains very much unperturbed (note that with increasing hole-pocket area, the self-consistency allows the associated coherence factors to adjust themselves to maintain the same doping concentrations). At some intermediate temperature, $T_{m}$, the `quasi-hole' and `quasi-electron' weights exactly cancel each other to make $R_H$ vanishes; see Fig.~\ref{fig1}(b). At $T>T_m$, the larger hole-pocket reverses the sign of the Hall-resistance, in Fig.~\ref{fig1}(e). Finally, above the SDW temperature, the metallic state is restored and the Hall coefficient becomes featureless. The same analysis applies to the higher doping result in that doping reduces the PG order, rendering a smaller hole-pocket, and hence less transport contribution. The results are in excellent agreement with the corresponding experimental data\cite{HallEP}, plotted with symbols in Fig.~\ref{fig2}(e) and (f).

The discussion of Nernst effect proceeds similarly, although the fundamental mechanism for the enhanced Nernst signal is somewhat different. Experimentally, it is well established that the Nernst signal changes sign from negative to positive as well as it becomes abruptly enhanced below the PG temperature, but much above the superconducting transition temperature, $T_c$,\cite{Nernst_nematic,Nernst_data} see inset to Fig.~\ref{fig2}(f). Despite some proposals that there exists vortex-type excitations even above $T_c$ which can drive Nernst signal\cite{Nernst_prepairing}, the widely used ambipolar phenomena suggests that the compensating electron-like and hole-like FSs are primarily necessary to obtain an enhanced `normal state' Nernst signal.\cite{ambipolar,Sachdev_SDW,DDW_Nernst} Within the present theory, such ambipolar state naturally evolves in the vicinity of the intermediate temperature $T_m$, where `quasi-hole' and `quasi-electron weights' become comparable, and a positive maximal of Nernst signal arises, as shown in Fig.~\ref{fig2}(b). Quantum-oscillation measurement has also detected similar compensating electron- and hole-like FSs in YBCO.\cite{sebastian_compensated} Away from this $T$ region, the positive Nernst signal diminishes as the underlying states depart from predominant ambipolar character, that is, when only one type of `quasi-charge weight' dominates the excitation spectrum. Interestingly, even when the metallic state appears above the onset temperature of the density wave order, the so-called Sondheimer cancelation of the Nernst signal in a metallic state is still not strictly applicable here due to the coexistence of two-carrier charges on different bands.\cite{Sondheimer} In fact, the finite experimental value of the Nernst signal above $T_s$, which remained an unexplained puzzle in the existing theories, is surprisingly well reproduced within the present theory. The results capture the salient experimental features of YBCO data available at a slightly different doping,\cite{Nernst_data,Nernst_data_foot} although some discrepancies are clearly visible. The adopted relaxation time approximation imposes limitation to quantitatively map out the detailed shape and peak positions of the Nernst signal. Furthermore, the present linear-response theory is mainly applicable to the low magnetic field region; on the other hand, experimentally it requires sufficiently high field to expose the `normal state' Nernst signal. Despite these approximations, the Hall and Nernst anomalies are robust features in YBCO, tied to the underlying electronic structure that is confirmed by ARPES\cite{ARPES_damascelli,ARPES_borisenko} and first-principle results\cite{LDA}.

{\it In-plane anisotropy:-}The added benefit of the present theory is that the existence of the the quasi-1D chain state naturally explains the associated in-plane anisotropy in the Nernst signal. For the 1D state lying along the (100)-direction, the corresponding quasiparticle velocity is $v_x\ll v_y$, which leads to a similar anisotropy in the Nernst conductivity $\hat{\alpha}$ defined in Eq.~\ref{alpha}, and hence in the Nernst coefficient $\nu/T$, plotted in Fig.~\ref{fig2}(f).  Most importantly, the in-plane anisotropy continues to persist even above the PG phenomena disappears, both in experiment and theory, which adds confidence to our conclusion that the PG phase in YBCO is unlikely to arise from any spontaneous rotational symmetry-breaking ordering. Furthermore, the metallic chain states have been demonstrated by ARPES data\cite{ARPES_damascelli} to be present for doping range as large as $p=0.05$ to $p=0.29$ (and above), and all the existing evidences of in-plane anisotropy via transport,\cite{transport_nematic,Nernst_nematic} neutron scattering measurement\cite{HinkovScience} are obtained within this doping range. This clearly indicates that the presence of chain state is responsible for the observed in-plane anisotropy in this system. It is worthwhile mentioning that most of the evidences for the electronic nematic phase in cuprate is obtained in YBCO family only,\cite{transport_nematic,Nernst_nematic,HinkovScience} with some indication of it is presented in an analysis of the STM data in Ba-based compound.\cite{EAKNature}

{\it Outlook and conclusion:-} To further support our postulates, we refer to several other experimental facts where the present mechanism can offer a consistent explanation, at least in principle. (1) An important finding of the quantum oscillation measurements is that the cyclotron frequency of the election-like FS--which is proportional to the quasiparticle density at $E_F$--is very much doping independent in the optimal doping region.\cite{sebastianmass,QOPDai} In this context, a phenomenological model can be recalled which has shown that quantum oscillation can arise from open-orbit `Fermi arc', by taking advantage of the electron-hole mixing of the Cooper pairs.\cite{refael} This argument finds support from an independent experimental consideration.\cite{open_orbit_QO} Based on this mechanism, the electron and hole quasiparticles, residing on different part of the FS in our present model, can also give rise to a quantum oscillation. If so, the doping independent cyclotron frequency arising from the weakly doping dependent chain state can naturally explain the experimental observation of the cyclotron mass. Another explanation follows in that the open-orbit chain state becomes `warped FS' and can create closed FS pockets if the orthorhombic-II unit cell is chosen, which within the conventional framework will give rise to quantum oscillation. (2) A second important result of recent quantum oscillations study is that the oscillation disappears at a critical value of the doping for YBCO, and the cyclotron mass diverges as the critical value is approached from the high doping side. Millis {\it et al.}\cite{Norman_Lifshitz} have argued that the mass divergence is related to a Lifshitz transition where the multiple FS pockets connect to form an open (quasi-1D) FS. Our argument follows similarly where we expect that in extreme underdoping, the CuO$_2$ states become vanishingly small, and the residual 1D chain states effectively cause a logarithmic divergence of the cyclotron mass. A rigorous calculation of the quantum oscillation would be of considerable benefit to confirm our predictions. (3) Finally, we see no fundamental error in postulating that the `electron-like' and `quasi-uniaxial' chain FS can also explain the negative Seebeck coefficient and thermopower signal,\cite{seebeck,seebeckNC} induced in-plane anisotropy in the spin-susceptibility measured by inelastic neutron scattering study.\cite{HinkovScience,DasINS} Finally, we mention that since the chain state remains mettalic in double chain layered YBa$_2$Cu$_4$O$_{8+\delta}$ (Y124),\cite{LDA} the above postulates continue to hold in this system, which is consistent with the observation of quantum oscillation in Y124.\cite{QOY124}

Our realistic framework narrows the range of possible theoretical models for the PG state in cuprates. It strongly suggests that the microscopic route toward understanding the nature of the PG physics may be generic to all cuprates containing a common CuO$_2$ plane. Most of the other accompanying normal state properties can then be explained as added features coming from the material specific crystallographic differences, such as apical oxygens surrounding Cu atoms on the CuO$_2$ planes in La- and Bi-based superconductors, chain layers in YBCO, chemically active Cl-ions in Ca$_{2-x}$Na$_x$CuO$_2$Cl$_2$, among others.

{\it Note added:} We argue that recent observations of charge density wave (CDW) do not alter our results. The reasons are: (1) The CDW has so far been observed in monolayer Bi-based cuprate\cite{Hudson} and YBCO\cite{fldCO,RIXS}, and its onset temperature and doping are very different from thePG temperature $T^*$. On the other hand, the psuedogap phenomena and similar FS properties are obtained in all hole doped cuprates. (2) A theoretical study\cite{Harrison} has shows that a biaxial CDW induced nodal FS pocket is electron-like. According to the definition of electron-pocket, it is associated with a band bottom or band folding below $E_F$, implying a gap opening along the nodal direction below $E_F$. But ARPES evidence for a gap along this line is yet far from definitive. Based on these facts, we argue that CDW is a secondary or weak material-specific feature, while the PG has more fundamental and generic origin.

\begin{acknowledgments}
The work is supported by the U.S. DOE through the Office of Science (BES) and the LDRD Program and benefited by NERSC computing allocation.
\end{acknowledgments}

\section{Supplementary Material}

\subsection{Eigenstates of the tri-layer Lattice model}

\begin{figure}
\hspace{-0cm}
\rotatebox{0}{\scalebox{.6}{\includegraphics{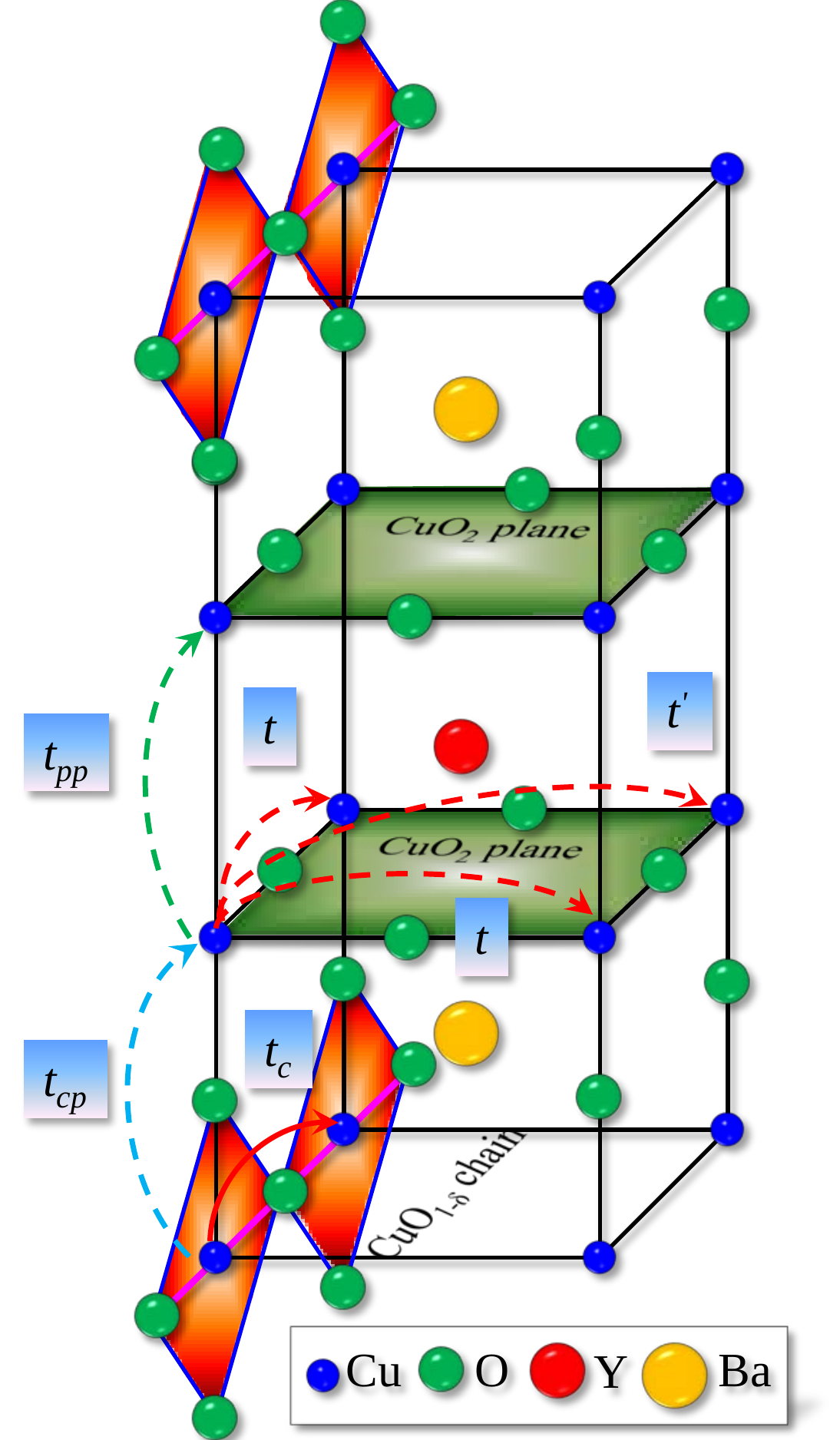}}}
\caption{Crystal structure and tight-binding hopping parameters of YBCU unit cell.}
\label{supp1}
\end{figure}

Supplementary Fig.~\ref{supp1} presents the unit cell of YBCO crystal. The unit-cell contains two chain-layers and two CuO$_2$ plane layers. In order to obtain a minimum low-energy model we assume that the inter-layer electron tunneling is only active between the the two nearest neighbor layers. In doing so, we are left with a tri-layer system which consists of two CuO$_2$ planes and one CuO chain, while the periodic boundary conditions on infinite lattice is imposed along all three dimensions. Furthermore, the chain state does not exhibit any gap opening through out the doping range of YBCO. The plane states undergo FS reconstruction which can be captured within a commensurate spin-density wave with modulation vector ${\bm Q}=(\pi,\pi)$. The commensurate modulation doubles the unit cell in real-space. Using the standard Nambu notation, we define the tri-layer eigenfunction in the magnetic zone as $\Psi_{\bm k}^{\dag}=\left[ c_{p{\bm k}\uparrow}^{\dag},~ c_{p^{\prime}{\bm k}\uparrow}^{\dag},~ c_{c{\bm k}\uparrow}^{\dag},~ c_{p({\bm k}+{\bm Q})\downarrow}^{\dag},~ c_{p^{\prime}({\bm k}+{\bm Q})\downarrow}^{\dag},~ c_{c({\bm k}+{\bm Q})\downarrow}^{\dag}\right]$. In this notation the Hamiltonian presented in the main text in Eq.~1 becomes a 6$\times$6 matrix $H = \sum_{\bm k}\Psi_{\bm k}^{\dag}H_{\bm k}\Psi_{\bm k}$:

\begin{widetext}
\begin{eqnarray}
H_{\bm{k}}=
\left(
\begin{array}{cccccc}\
\xi_{pp{\bm k}} & \xi_{pp^{\prime}{\bm k}} & \xi_{cp{\bm k}} & V_{pp{\bm k}} & 0 & 0 \\
\xi_{pp^{\prime}{\bm k}} & \xi_{pp{\bm k}} & \xi_{cp{\bm k}} & 0 & V_{p^{\prime}p^{\prime}{\bm k}} & 0\\
\xi_{cp{\bm k}} & \xi_{cp{\bm k}} & \xi_{cc{\bm k}} & 0 & 0 & 0\\
V_{pp{\bm k}} & 0 & 0 & \xi_{pp({\bm k}+{\bm Q})} & \xi_{pp^{\prime}({\bm k}+{\bm Q})} & \xi_{cp({\bm k}+{\bm Q})}\\
0 & V_{p^{\prime}p^{\prime}{\bm k}} & 0 & \xi_{pp^{\prime}({\bm k}+{\bm Q})} & \xi_{pp({\bm k}+{\bm Q})} & \xi_{cp({\bm k}+{\bm Q})}\\
0 & 0 & 0 & \xi_{cp({\bm k}+{\bm Q})} & \xi_{cp({\bm k}+{\bm Q})} & \xi_{cc({\bm k}+{\bm Q})}\\
\end{array}
\right).\label{Hamk}
\end{eqnarray}
\end{widetext}

In obtaining the non-interacting dispersions $\xi_{ij{\bm k}}$, we assume that Cu $d_{x^2-y^2}$ orbital contributes to the low-energy scales of present interest (as commonly used for all cuprates). The Cu $d_{x^2-y^2}$ electrons hop to their neighbors via O $p_{x,y}$ orbitals which can be taken into account within effective tight-binding formalism.\cite{markietb,atkinson} The hoping terms are depicted in supplementary Fig.~\ref{supp1}, which gives,
\begin{eqnarray}
&&\xi_{pp{\bm k}} =-2t(\phi_x+\phi_y) -4t^{\prime}\phi_x \phi_y -2t^{\prime\prime}(\phi_{2x}+\phi_{2y})-\mu_p,\nonumber\\
&&\xi_{cc{\bm k}} =-2t_{c}\phi_y-\mu_c, \nonumber\\
&&\xi_{pp^{\prime}{\bm k}} =-2t_{pp^{\prime}}(\phi_x-\phi_y)\phi_{z/4},~~~\xi_{cp} =-2t_{cp}\phi_{z/4},
\label{disp}
\end{eqnarray}
where $\phi_{\alpha \mu}$=$\cos{(\alpha k_{\mu})}$ with $\mu$=$x$,$y$,$z$. $\mu_{p,c}$ are the chemical potential for the plane and chain states, respectively, which encode relative onsite energy differences and other crystal effects between the two levels. $t$,$t^{\prime}$, and $t^{\prime\prime}$ are the 1st, 2nd and 3rd nearest neighbor (NN) hopping on the 2D CuO$_2$ plane, $t_c$ is the NN hopping on the 1D chain aligned along $y$-axis, and $t_{pp^{\prime}}~(t_{cp})$ is the plane-plane (chain-plane) hopping along $z-$direction.

The eigenvalues for the upper 3$\times$3 interblock matrix is obtained as
\begin{eqnarray}
E_{C{\bm k}}&=&\epsilon^+_{\bm k} + E_{0{\bm k}}\\
E_{A{\bm k}}&=&\epsilon^+_{\bm k} - E_{0{\bm k}}\\
E_{B{\bm k}}&=&\chi^-_{\bm k}.
\end{eqnarray}
Here subscripts `A' and `B' stand for the anti-bonding and bonding states created by the bi-CuO$_2$ layers, where `C' denotes the chain state. $\epsilon^{\pm}_{\bm k} = (\chi^+_{\bm k}\pm\xi_{cc{\bm k}})/2$, $\chi^{\pm}_{\bm k}=\xi_{pp{\bm k}} \pm \xi_{pp^{\prime}{\bm k}}$, and $E_{0{\bm k}} = \sqrt{((\epsilon^{-}_{\bm k})^2 + 2\xi^2_{cp{\bm k}}}$. Of course, the weight of the each layers are mixed between each bands which can be determined by their corresponding eigenvectors:
%
%\begin{widetext}
%\begin{eqnarray}
%\psi_{C{\bm{k}}}=
%\frac{1}{\sqrt{3}}\left(
%\centering
%\begin{array}{c}\
%u^{+}_{\bm k} \\
%
%u^{-}_{\bm k} \\
%%
%1
%\end{array}
%\right),~
%%
%\psi_{A{\bm{k}}}=
%\frac{1}{\sqrt{3}}\left(
%\centering
%\begin{array}{c}\
%u^{-}_{\bm k} \\
%%
%u^{+}_{\bm k} \\
%%
%1
%\end{array}
%\right),~{\rm and}~
%%
%\psi_{B{\bm{k}}}=
%\frac{1}{\sqrt{3}}\left(
%\centering
%\begin{array}{c}\
%1\\
%1\\
%1
%\end{array}
%\right),
%%.\label{Ham}
%\end{eqnarray}
%%
%\end{widetext}
%where 
$u^{\pm}_{\bm k} = \frac{1}{2}\left[\epsilon^{-}_{\bm k}\pm E_{0{\bm k}}\right]/{\xi_{cp{\bm k}}}$. The eigenvalues of the full Hamiltonian can now be expressed as two split bands of the above three bands gapped by SDW order parameter $V_{ij}$\cite{DasSDW,DDW}:
\begin{eqnarray}
E^{\pm}_{C{\bm k}}&=&\xi^+_{C{\bm k}} \pm \xi^-_{C{\bm k}} \\
\label{E06}
E^{\pm}_{A{\bm k}}&=&\xi^+_{A{\bm k}} \pm \sqrt{\left(^-\xi_{A{\bm k}}\right)^2 + V_A^2} \\
\label{E16}
E^{\pm}_{B{\bm k}}&=&\xi^+_{B{\bm k}} \pm \sqrt{\left(\xi^-_{B{\bm k}}\right)^2 + V_B^2},
\label{E6}
\end{eqnarray}
where $\xi^{\pm}_{i{\bm k}}=\left(E_{A{\bm k}}\pm E_{A{\bm k}}\right)/2$ and the self-consistent order parameters $V_i$ are given below. The eigenvectors for the final eigenstates given above can be represented in terms of SDW coherence factors:
\begin{eqnarray}
\alpha^2_{i{\bm k}} &=&\frac{1}{2}\left( 1+ \xi^-_{i{\bm k}}/\sqrt{\left(\xi^-_{i{\bm k}}\right)^2 + V_i^2}\right),\nonumber\\
\beta^2_{i{\bm k}} &=& 1-\alpha^2_{i{\bm k}}.
\end{eqnarray}
Here $i=A,B,C$. Due to the absence of any gap opening in the chain state, $\alpha^2_{C{\bm k}}=1$ and $\beta^2_{C{\bm k}}=0$ at all momentum. With the definitions of the eigenstates and the coherence factors, we construct the final unitary matrix $\hat{U}_{\bm k}$ that ultimately diagonalize the total Hamiltoanin given above in Eq.~\label{Hamk}
%

%\begin{widetext}
\begin{eqnarray}
{\hat U}_{\bm{k}}=
\left(
\begin{array}{cccccc}\
u^+_{\bm k} & u^-_{\bm k}\alpha_{A{\bm k}} & -\alpha_{B{\bm k}} & 0 & -u^-_{\bm k}\beta_{A{\bm k}} & \beta_{B{\bm k}}  \\
u^-_{\bm k}\alpha_{A{\bm k}} & u^+_{\bm k}\alpha_{A{\bm k}} & \alpha_{B{\bm k}} & 0 & -u^+_{\bm k}\beta_{A{\bm k}} & -\beta_{B{\bm k}}\\
-\alpha_{B{\bm k}} & \alpha_{B{\bm k}} & 0 & 0 & -\beta_{B{\bm k}} & 0\\
0 & 0 & 0 & 0 & 0 &0 \\
u^-_{\bm k}\beta_{A{\bm k}} & u^+_{\bm k}\beta_{A{\bm k}} & \beta_{B{\bm k}} & 0 & u^+_{\bm k}\alpha_{A{\bm k}} & \alpha_{B{\bm k}}\\
-\beta_{B{\bm k}} & \beta_{B{\bm k}} & 0 & 0 & \alpha_{B{\bm k}} & 0
\end{array}
\right).\nonumber\\
\label{Uk}
\end{eqnarray}
%
%\end{widetext}

The self-consistent order parameters are defined as $V_A=US_A$ and $V_B=US_B$ (and $V_C=0$), where $U$ is the onsite Hubbard interaction which is kept to be same for both CuO$_2$ planes. $S_{A/B}$ are the two order parameters, $p$ is the dopin concentration which are evaluated by solving the following two self-consistent equations:
\begin{eqnarray}
p &=& 2 - \frac{1}{N}\sum_{{\bm k},i=1,2,\sigma} \left\langle c_{i{\bm k}\sigma}^{\dag}c_{i{\bm k}\sigma}\right\rangle\nonumber\\
&=&2-\frac{1}{N}\sum_{{\bm k},i}\hat{U}_{ii{\bm k}}\hat{U}_{ii{\bm k}}^{\dag}f(E_{i{\bm k}}),\\
S_{i} &=&\frac{1}{2}\left(n_{i\uparrow}-n_{i\downarrow}\right)\nonumber\\
&=&\frac{1}{2N}\sum_{\bm k}\left[\left\langle c_{i({\bm k}+{\bm Q})\uparrow}^{\dag}c_{i{\bm k}\uparrow}\right\rangle -\left\langle c_{i({\bm k}+{\bm Q})\downarrow}^{\dag}c_{i{\bm k}\downarrow}\right\rangle \right]\nonumber\\
&=&\frac{1}{2N}\sum_{{\bm k},j}\hat{U}_{ij{\bm k}}\hat{U}_{ji{\bm k}}^{\dag}f(E_{j{\bm k}}),
\end{eqnarray}
where $f$ is the Fermi function and $E_j$ are the eigenstates listed in Eq.~\ref{E06}-\ref{E6} where indices $i$=A,B and $j$=1-6. $n_{i\uparrow}$ is the number of up-spin on the $i^{th}$ layer defined as $\left\langle c_{i({\bm k}+{\bm Q})\uparrow}^{\dag}c_{i{\bm k}\uparrow}\right\rangle$, where the thermal average is taken over the whole Brillouin zone. 

\vspace{12pt}
{\bf Tight-binding parameters and self-consistent order parameters}

\begin{figure}
\hspace{-0cm}
\rotatebox{0}{\scalebox{.6}{\includegraphics{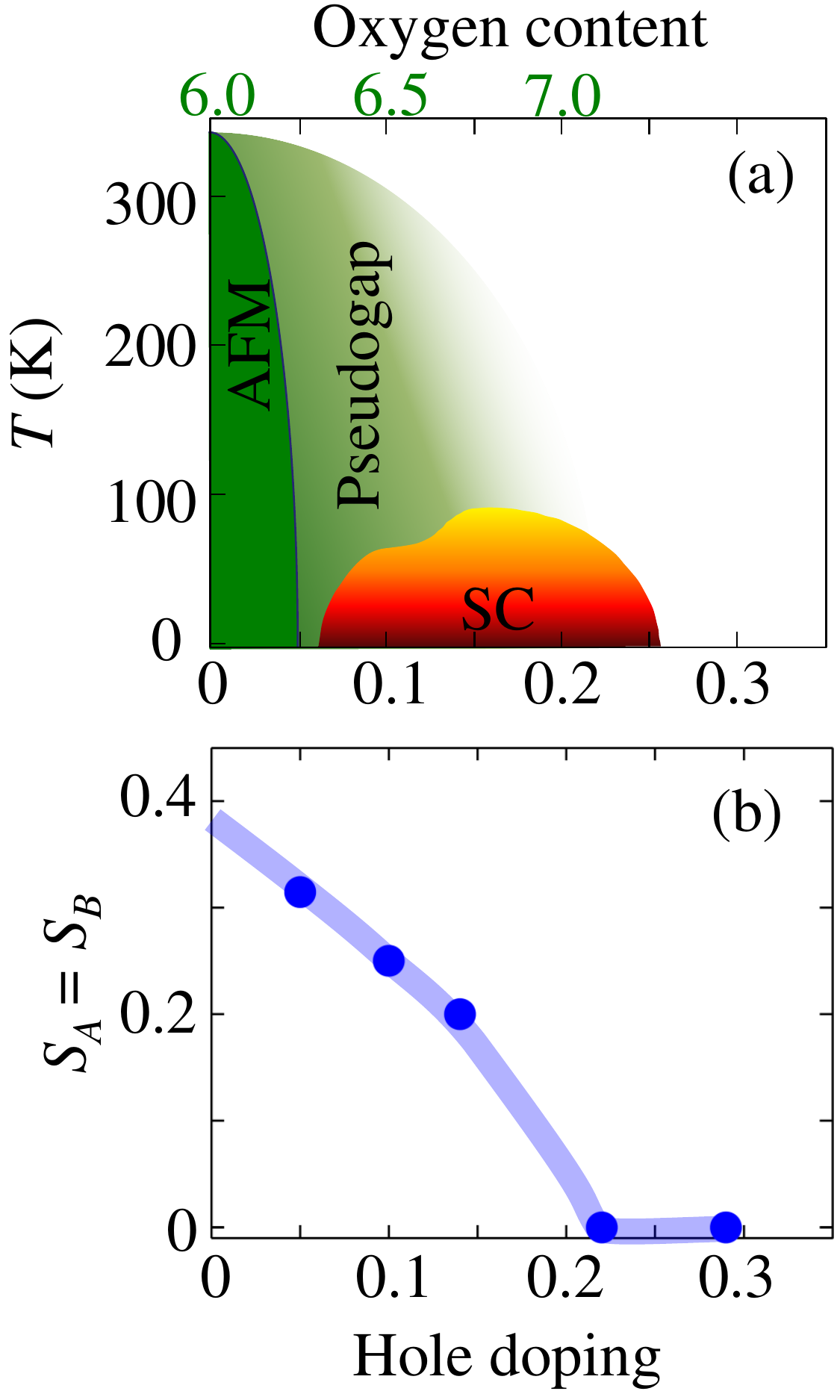}}}
\caption{(a) Schematic phase diagram of cuprate.  (b) The doping dependence of self-consistently evaluated order parameter $S$ (defined in the text).}
\label{supp2}
\end{figure}

We find that the tight-binding parameters that give a good description of the FS, in agreement with ARPES and first-principle band-structure are\cite{atkinson}: $(t,t^{\prime},t^{\prime\prime},t^{\prime\prime},t_c,t_{pp^{\prime}},t_{cp})$=(0.38, 0.0684,0.095,0.25,-0.01,-0.0075) eV. We have kept $\mu_c=-0.87$~eV to be doping independent, while $\mu_p$ is computed self-consistently. The band structure fitting is done at the overdoped sample of $p=0.29$, shown in Fig.~1 in the main text.

The calculation of self-consistency is carried out in the following steps. At a given doping $p$ and onsite Hubbard $U$=1.59 eV, we start our self-consistent cycle with an initial guess of chemical potential $\mu_p$ and order parameters $S_{A/B}$, and calculate doping $p$ and $S_{A/B}$ by solving Eqs. (1) to (12) at $T=0$.  We iterate the entire loop until the self-consistent values of doping $p$, and order parameters $S_{A/B}$ converges. For the overdoped sample of $p=0.29$, we get $S_{A/B}=0$ and $\mu_p$=-0.54~eV. At an extreme underdoped samples of $p=0.05$ at which ARPES data are available for fitting in Fig.~1, we get $S_A(T=0)\approx S_B(T=0)=0.3145$ and $\mu_p=-0.4$. At two dopings at which Fig.~2 is obtained in the main text, we find $S_A(T=0)\approx S_B(T=0)=0.25$ and $\mu_p$=-0.41 at $T$=0 to -0.43~eV at $T=55$~K for $p$=0.10, and $S_A(T=0)\approx S_B(T=0)=0.22$ and $\mu_p$=-0.38 at $T$=0 to -0.41~eV at $T=55$~K for $p$=0.14. The results are plotted in supplementary Fig.~\ref{supp2}(b), and compared with the schematic phase diagram of YBCO in Fig.~1(a). Note that the temperature dependence of the order parameters is phenomenologically assumed as $S_i=S_i(0)\sqrt{1-T/T_{s}}$, where $T_s=55$~K is the SDW transition temperature.\cite{Sachdev_stripe,Ts}

%
%\vspace{12pt}
\section{Details of Hall and Nernst coefficient calculations.}

The Nernst effect experiments measure the transverse electric field response of a system to a combination set-up of an externally-imposed temperature gradient and an orthogonal magnetic field. The Nernst effect depends sensitively on anisotropies in the band structure. Similarly, the Hall effect is a powerful probe of the Fermi surface of a metal because of its sensitivity to the sign of charge carriers, which distinguishes between electrons and holes. These two probes have recently gained much attention in the context of cuprates, since they demonstrate very unusual behavior in the underdoped region when the pseudogap sets in. Apart from the superconducting fluctuation origin of enhanced Nernst effect in cuprates,\cite{Nernst_prepairing} purely quasiparticle picture is extensively proposed in various contexts.\cite{ambipolar,Sachdev_SDW,Sachdev_stripe,DDW_Nernst} The quasiparticle Nernst effect has been studied on the basis of the linearized Boltzmann equation in the relaxation-time approximation. This is a reliable approach if the scattering rate is isotropic, since then the neglected `scattering-in' contributions average out to zero.

We start from the semiclassical Boltzmann transport equation for quasiparticles (charge $e$) in a weak magnetic field ${\bm B}$, driven out of equilibrium by a spatially uniform electric field ${\bm E}$ and temperature gradient $\nabla T$. If we linearize the deviation of the fermion distribution function from its equilibrium state $f(\xi_{\bm k})$ in both ${\bm E}$ and $\nabla T$ as $f'(\xi_{\bm k})=f + g_{\bm k}$, then the Boltzmann equation\cite{Nernst_QSH} becomes:
\begin{eqnarray}
&&V_{\bm k}\left(-\frac{\partial f(\xi_{\bm k})}{\partial \xi_{\bm k}}\right) - \Omega^0_{\bm k}g_{\bm k} = \sum_{{\bm k}^{\prime}} Q_{{\bm k},{\bm k}^{\prime}}\left(g_{\bm k}-g_{{\bm k}^{\prime}}\right)
\label{eqn10}\\
&&V_{\bm k}= {\bm v}_{\bm k}\cdot\left(-e{\bm E}-\xi_{\bm k}\frac{\nabla T}{T}\right)
\label{eqn11}\\
&&\Omega_{\bm k}=-\frac{e}{\hbar c}\left({\bm v}_{\bm k}\times{\bm B}\right)\cdot\nabla_{\bm k}.
\label{eqn12}
\end{eqnarray}
Here $\xi_{\bm k}$ is the quasiparticle state whose velocity is ${\bm v}_{\bm k}=\hbar^{-1}\nabla_{\bm k}\xi_{\bm k}$. The band index and the eigenvectors are implicit to the quasiparticle states in the above equations. $Q_{{\bm k},{\bm k}^{\prime}}$ is the scattering matrix-element for incoming state $\sum_{{\bm k}^{\prime}}Q_{{\bm k},{\bm k}^{\prime}}g_{{\bm k}^{\prime}}$ and outgoing state $\sum_{{\bm k}^{\prime}}Q_{{\bm k},{\bm k}^{\prime}}g_{{\bm k}}$. For the present DC transport property, we assume the elastic scattering with $Q_{{\bm k},{\bm k}^{\prime}}=\delta(\xi_{\bm k}-\xi_{{\bm k}^{\prime}})q_{{\bm k},{\bm k}^{\prime}}$. Furthermore, we impose the standard relaxation-time approximation,\cite{RelaxA} which implies that the system has enough time to loose its memory, i. e. scattering-in term is negligible. In that case, the scattering rate is governed
is entirely governed by the out-going state as $\frac{g_{\bm k}}{\tau_{\bm k}}=g_{\bm k}\sum_{{\bm k}^{\prime}}Q_{{\bm k},{\bm k}^{\prime}}$, while relating its momentum dependence:
\begin{eqnarray}
\frac{1}{\tau_{\bm k}}=\sum_{{\bm k}^{\prime}}{\hat Q}_{{\bm k},{\bm k}^{\prime}}=N_{0}\left\langle q_{{\bm k},{\bm k}^{\prime}}\right\rangle_{FS}.
\label{tau}
\end{eqnarray}

Within this approximation Eq.~\ref{eqn10}-\ref{eqn12} becomes

\begin{eqnarray}
&&g_{\bm k}=\Omega_{\bm k}^{-1}V_{\bm k}\left(-\frac{\partial f(\xi_{\bm k})}{\partial \xi_{\bm k}}\right)\\
\label{eqn20}
&&V_{\bm k}= {\bm v}_{\bm k}\cdot\left(-e{\bm E}-\xi_{\bm k}\frac{\nabla T}{T}\right)
\label{eqn21}\\
&&\Omega_{\bm k}=-\left[\frac{e}{\hbar c}\left({\bm v}_{\bm k}\times{\bm B}\right)\cdot\nabla_{\bm k} + \frac{1}{\tau_{\bm k}}\right].
\label{eqn22}
\end{eqnarray}
It is now convenient to introduce the band index $n$:
%To do so, we project the non-equilibrium distribution operator from its present band basis $n$ (a band index is implicit on both sides of Eq.~\ref{eqn21}) to the layer indices, $i,j$ as
%
%
\begin{eqnarray}
&&g_{n{\bm k}}=\Omega_{n\bm k}^{-1}V_{n\bm k}\left(-\frac{\partial f(\xi_{n\bm k})}{\partial \xi_{n\bm k}}\right),
\label{eqn31}
%
%&&M_{n}^{ij}({\bm k})= \psi_n^i({\bm k})\psi_n^{j\dag}({\bm k}).
\end{eqnarray}
%
%where $M$ is the matrix-element.
From Eq.~\ref{eqn31}, the net electrical, ${\bm J}$, and thermal current densities, ${\bm Q}$, that flow on the $i^th$-layer (or orbital) can be calculated as from the diagonal term of the following matrix
\begin{eqnarray}
{\bm J}^{ij}&=&-2e\sum_{{\bm k},n}M_{n}^{ij}{\bm v}_{n\bm k}g_{n\bm k}
\label{eqn41}\\
{\bm Q}^{ij}&=&2\sum_{{\bm k},n}M_{n}^{ij}{\bm v}_{n\bm k}\xi_{n{\bm k}}g_{n\bm k},
\label{eqn51}
\end{eqnarray}
where $M$ is the matrix-element:
\begin{eqnarray}
M_{n}^{ij}({\bm k})= \psi_n^i({\bm k})\psi_n^{j\dag}({\bm k}).
\end{eqnarray}
In Eq.~\ref{eqn41}, we project the densities from its band basis $n$ (a band index is implicit on both sides of Eq.~\ref{eqn20}) to the layer indices, $i,j$. The factor `2' introduced in Eqs.~\ref{eqn41}-\ref{eqn51} to count spin-degeneracy.

{\bf Linear response theory:} The interplay of electrical and thermal effects necessarily implies two conductivity tensors $\hat{\sigma}$,  $\hat{\alpha}$, which relate charge current ${\bm J}$ to electric field, ${\bm E}$ and thermal gradient, $\nabla T$ vectors:

\begin{eqnarray}
{\bm J}^{ij} = \hat{\sigma}^{ij}{\bm E}-\hat{\alpha}^{ij}\nabla T
%
%\left(
%\begin{array}{c}\
%{\bm J} \\
%{\bm Q}
%\end{array}
%\right)=
%%
%\left(
%\begin{array}{cc}\
%\hat{\sigma} & \hat{\alpha} \\
%T\hat{\sigma} & \hat{\kappa} \\
%\end{array}
%\right)
%
%\left(
%\begin{array}{c}\
%{\bm E} \\
%-\nabla T
%\end{array}
%\right).
%
\label{mat}
\end{eqnarray}
The Nernst response is defined as the electricalfield induced by a thermal gradient in the absence of an electrical current, and is given in linear response by the relation ${\bm E}=-\hat{\theta}\nabla T$. In absence of charge current (i.e. when ${\bm J}$=0), Eq. (6) yields:
\begin{equation}
{\bm E}=(\hat{\sigma}^{ij})^{-1}\hat{\alpha}^{ij}\nabla T=-\hat{\theta}^{ij}\nabla T,
\label{ET}
\end{equation}
where Hall resistance, $\hat{\rho}$=$\hat{\sigma}^{-1}$, and the Nernst coefficient, ${\nu}$=$\hat{\theta}/B$, are measured independently via transport probes by setting a weak magnetic field ${\bm B}$=$B\hat{z}$ along the $c$-axis of the lattice.\cite{HallEP,Nernst_data} In Eq.~\ref{mat} and \ref{ET}, the layer indices $i,j$ run on both sides. From Eqs.~\ref{eqn21},\ref{eqn41}, \ref{mat}, we easily deduce that
\begin{widetext}
\begin{eqnarray}
\hat{\sigma}^{ij}_{\mu\nu}&= &2\sum_{{\bm k},n}M_{n}^{ij}{\bm v}^{\mu}_{n\bm k}\Omega_{n\bm k}^{-1}\left(\frac{\partial V^{\nu}_{n\bm k}}{\partial {\bm E}}\right)\left(-\frac{\partial f(\xi_{n\bm k})}{\partial \xi_{n\bm k}}\right)
=\frac{\beta e^2}{2}\sum_{{\bm k},n}M_{n{\bm k}}^{ij}v^{\mu}_{n{\bm k}}\Omega_{n{\bm k}}^{-1}v^{\nu}_{n{\bm k}}{\rm sech}^2\left(\frac{\beta E_{n{\bm k}}}{2}\right)\\
\hat{\alpha}^{ij}_{\mu\nu}&=& 2\sum_{{\bm k},n}M_{n}^{ij}{\bm v}^{\mu}_{n\bm k}\Omega_{n\bm k}^{-1}\left(\frac{\partial V^{\nu}_{n\bm k}}{\partial (-\nabla T)}\right)\left(-\frac{\partial f(\xi_{n\bm k})}{\partial \xi_{n\bm k}}\right)=-\frac{\beta^2 e}{2}\sum_{{\bm k},n}M_{n{\bm k}}^{ij}v^{\mu}_{n{\bm k}}\Omega_{n{\bm k}}^{-1}v^{\nu}_{n{\bm k}}E_{n{\bm k}}{\rm sech}^2\left(\frac{\beta E_{n{\bm k}}}{2}\right)
\label{sigalp2}
\end{eqnarray}
\end{widetext}
where we have used the identity $\left(-\frac{\partial f(\xi_{n\bm k})}{\partial \xi_{n\bm k}}\right)=\frac{\beta}{4}{\rm sech}^2\left(\frac{\beta E_{n{\bm k}}}{2}\right)$. In the usual manner, the differential operator $\Omega_{n\bm k}^{-1}$ can be arranged as a perturbative expansion in the magnetic field ${\bm B}$\cite{Sachdev_stripe,RelaxA} in order to obtain transport coefficients that do not depend on ${\bm B}$. For this purpose we define  $\Omega_{n\bm k}$= $\Upsilon_{n\bm k}$+$\Lambda_{n\bm k}$ (Eq.~\ref{eqn12}), where $\Upsilon_{n\bm k}=1/\tau_{n\bm k}$ and $\Lambda_{n\bm k}$ is the rest. Then
\begin{eqnarray}
\Omega_{n\bm k}^{-1}=\Upsilon_{n\bm k}^{-1} -\Upsilon_{n\bm k}^{-1}\Lambda_{n\bm k}\Upsilon_{n\bm k}^{-1} + \mathcal{O}(B^2).
\label{sigalp3}
\end{eqnarray}
The diagonal entries in Eq.~\ref{sigalp2} (in the $\mu,\nu$ basis, not in the $i,j$ basis) are obtained from the zeroth order in ${\bm B}$ in Eq.~\ref{sigalp3}, while the lowest-order contribution to the off-diagonal coefficients arises from the linear order in ${\bm B}$ in the expansion of Eq.~\ref{sigalp3}. To this accuracy, the expressions Eq.~\ref{sigalp2} can be simplified in form of
the expressions
\begin{widetext}
\begin{eqnarray}
\hat{\sigma}^{ij}_{\mu\mu}&=&\frac{\beta e^2}{2}\sum_{{\bm k},n}M_{n{\bm k}}^{ij}(v^{\mu}_{n{\bm k}})^2\tau_0{\rm sech}^2\left(\frac{\beta E_{n{\bm k}}}{2}\right)\\
\hat{\sigma}^{ij}_{\mu\ne\nu}&=&\frac{\beta e^3B}{2\hbar c}\sum_{{\bm k},n}M_{n{\bm k}}^{ij}v^{\mu}_{n{\bm k}}\tau_0^2\left[v^{\nu}_{n{\bm k}}\frac{\partial v^{\nu}_{n{\bm k}}}{\partial {\bm k}^{\mu}}-v^{\mu}_{n{\bm k}}\frac{\partial v^{\nu}_{n{\bm k}}}{\partial {\bm k}^{\nu}}\right]{\rm sech}^2\left(\frac{\beta E_{n{\bm k}}}{2}\right)\\
\hat{\alpha}^{ij}_{\mu\mu}&=& -\frac{\beta^2 e}{2}\sum_{{\bm k},n}M_{n{\bm k}}^{ij}(v^{\mu}_{n{\bm k}})^2\tau_0E_{n{\bm k}}{\rm sech}^2\left(\frac{\beta E_{n{\bm k}}}{2}\right)\\
\hat{\alpha}^{ij}_{\mu\ne\nu}&=& -\frac{\beta^2 e^2B}{2\hbar c}\sum_{{\bm k},n}M_{n{\bm k}}^{ij}v^{\mu}_{n{\bm k}}\tau^2_0\left[v^{\nu}_{n{\bm k}}\frac{\partial v^{\nu}_{n{\bm k}}}{\partial {\bm k}^{\mu}}-v^{\mu}_{n{\bm k}}\frac{\partial v^{\nu}_{n{\bm k}}}{\partial {\bm k}^{\nu}}\right]E_{n{\bm k}}{\rm sech}^2\left(\frac{\beta E_{n{\bm k}}}{2}\right)
\label{sigalp4}
\end{eqnarray}
\end{widetext}

In a simplest model, we assume a momentum and temperature independent value of relaxation time $\tau_{\bm k}=\tau_0$. A temperature dependent value of $\tau_0$ can improve the results, especially for the Nernst signal, however, with this simplest form, we are able to capture the salient features of both Hall coefficient and Nernst coefficient. The results in the main text are normalized to the constant value of $B$ and $\tau_0$. The above expressions matches exactly with the one proposed earlier for various form of density wave order,\cite{Sachdev_stripe,Sachdev_SDW,DDW_Nernst} except the matrix-element terms. It is important to note that, if no other crystal symmetry is broken, then the Mott relation $\nu_{xy}=- \nu_{yx}$ for Nernst coefficient can be recovered from the Boltzmann theory.\cite{Vojta} This implies that the choice of sign convention for the Nernst effect is arbitrary. To faciliate direct comparison, we have taken the same sign for $|\nu_a|$ and $|\nu_b|$.


\begin{thebibliography}{99}
\bibitem{HallEP}D. LeBoeuf {\it et al.},
%LeBoeuf, D. {\it et al.}
%Electron pockets in the Fermi surface of hole-doped high-$T_c$ superconductors.
{Nature} {\bf 450}, 533 (2007).
%David LeBoeuf1, Nicolas Doiron-Leyraud1, Julien Levallois2, R. Daou1, J.-B. Bonnemaison1, N. E. Hussey3, L. Balicas4, B. J. Ramshaw5, Ruixing Liang5,6, D. A. Bonn5,6, W. N. Hardy5,6, S. Adachi7, Cyril Proust2 & Louis Taillefer
%
\bibitem{seebeck} J. Chang {\it et al.},
%Chang, J. {\it et al.}
%Nernst and Seebeck coefficients of the cuprate superconductor YBa$_2$Cu$_3$O$_{6.67}$: A study of Fermi surface reconstruction.
{Phys. Rev. Lett.} {\bf 104}, 057005 (2009).
%R. Daou, Cyril Proust, David LeBoeuf, Nicolas Doiron-Leyraud, Francis Laliberte, B. Pingault, B. J. Ramshaw, Ruixing Liang, D. A. Bonn, W. N. Hardy, H. Takagi, A. Antunes, I. Sheikin, K. Behnia, Louis Taillefer
%
\bibitem{seebeckNC}F. Lalibert\"e {\it et al.},
%Lalibert\"e, F. {\it et al.}
%Fermi-surface reconstruction by stripe order in cuprate superconductors,
{Nat. Comm.} {\bf 2}, 432 (2011).
%F. Lalibert1, J. Chang1, N. Doiron-Leyraud1, E. Hassinger1, R. Daou, M. Rondeau1, B.J. Ramshaw2, R. Liang2,3, D.A. Bonn2,3, W.Hardy2,3, S. Pyon4, T. Takayama4, H. Takagi4,5, I. Sheikin6, L. Malone7, C. Proust3,7, K. Behnia8 & Louis Taillefer
%
\bibitem{Nernst_nematic}R. Daou {\it et al.},
%Broken rotational symmetry in the PG phase of a high-$T_c$ superconductor.
{Nature} {\bf 463}, 519 (2010).
%Daou, R. {\it et al.} Broken rotational symmetry in the PG phase of a high-$T_c$ superconductor. {\it Nature} {\bf 463}, 519 (2010).
%R. Daou1{, J. Chang1, David LeBoeuf1, Olivier Cyr-Choinie`re1, Francis Laliberte, Nicolas Doiron-Leyraud1, B. J. Ramshaw2, Ruixing Liang2,3, D. A. Bonn2,3, W. N. Hardy2,3 & Louis Taillefer1,3
%
\bibitem{Nernst_data}J. Chang {\it et al.},
%Chang, J. {\it et al.}
%Nernst effect in the cuprate superconductor YBCO: Broken rotational and translational symmetries.
{Phys. Rev. B} {\bf 84}, 014507 (2011).
%J. Chang, Nicolas Doiron-Leyraud, Francis Lalibert, R. Daou, David LeBoeuf, B. J. Ramshaw, Ruixing Liang, D. A. Bonn, W. N. Hardy, Cyril Proust, I. Sheikin, K. Behnia, Louis Taillefer
%
\bibitem{SdH2007}N. Doiron-Leyraud {\it et al.},
%Doiron-Leyraud, N. {\it et al.}
%Quantum oscillations and the Fermi surface in an underdoped high-$T_c$ superconductor.
{Nature} {\bf 447}, 565 (2007).
%
\bibitem{SdH2008}A. F. Bangura {\it et al.}, Phys. Rev. Lett. 100, 047004 (2008).
%, J. D. Fletcher, A. Carrington, J. Levallois, M. Nardone, B. Vignolle, P. J. Heard, N. Doiron-Leyraud, D. LeBoeuf, L. Taillefer, S. Adachi, C. Proust, and N. E. Hussey 
%
\bibitem{sebastian_compensated}S. E. Sebastian {\it et al.},
%Sebastian, S. E. {\it et al.} Compensated electron and hole pockets in an underdoped high-$T_c$ superconductor.
{Phys. Rev. B} {\bf 81}, 214524 (2010).
%Suchitra E. Sebastian, N. Harrison, P. A. Goddard, M. M. Altarawneh, C. H. Mielke, Ruixing Liang, D. A. Bonn, W. N. Hardy, O. K. Andersen, G. G. Lonzarich
%
\bibitem{sebastianmass}S. E. Sebastian {\it et al.},
%Sebastian, S.E. {\it et al.} Metal-insulator quantum critical point beneath the high-$T_c$ superconducting dome.
{Proc. Nat. Acad. Sci. USA} {\bf 107}, 6175 (2010).
% N. Harrison, M. M. Altarawneh, C. H. Mielke, Ruixing Liang, D. A. Bonn, W. N. Hardy, G. G. Lonzarich
%
\bibitem{QOPDai}J. Singleton {\it et al.},
%Singleton, J. {\it et al.}
%Magnetic quantum oscillations in YBa$_2$Cu$_3$O$_{6.61}$ and YBa$_2$Cu$_3$O$_{6.69}$ in fields of up to 85 T; patching the hole in the roof of the superconducting dome.
{Phys. Rev. Lett.} {\bf 104}, 86403 (2010).
%John Singleton, Clarina de la Cruz, R.D. McDonald, Shiliang Li, Moaz Altarawneh, Paul Goddard, Isabel Franke, Dwight Rickel, C.H. Mielke, Xin Yao, Pengcheng Dai
%
\bibitem{Onsagar} L.  Onsager,
%Onsager, L. Interpretation of the de Haas-van Alphen effect.
{Phil. Mag.} {\bf 43}, 1006 (1952);
%
\bibitem{ARPES_damascelli}D. Fournier {\it et al.},
%Loss of nodal quasiparticle integrity in underdoped YBa$_2$Cu$_3$O$_{6+x}$.
{Nat. Phys.} {\bf 6}, 905 (2010).
 %G. Levy, Y. Pennec, J.L. McChesney, A. Bostwick, E. Rotenberg, R. Liang, W.N. Hardy, D.A. Bonn, I.S. Elfimov, A. Damascelli
%
\bibitem{LDA}A. Carrington, and E. A. Yelland,
% Fermi surface pockets in ortho-II YBa$_2$Cu$_3$O$_{6.5}$: the origin of quantum oscillations?
{Phys. Rev. B} {\bf 76}, 140508(R) (2007);
%
\bibitem{Norman}A. J. Millis, and M. R. Norman,
% Antiphase stripe order as the origin of electron pockets observed in 1/8-hole-doped cuprates.
{Phys. Rev. B} {\bf 76}, 220503 (2007).
%
\bibitem{Norman_Lifshitz}A. J. Millis, and M. R. Norman,
%Lifshitz transition in underdoped cuprates.
{Phys. Rev. B} {\bf 81}, 180513(R) (2010).
%
\bibitem{DDW}S. Chakravarty, and H. Y. Kee,
%Fermi pockets and quantum oscillations of the Hall coefficient in high-temperature superconductors.
{Proc. Nat. Acad. Sci. USA} {\bf 105}, 8835 (2008).
%
\bibitem{DasSDW}T. Das, R. S. Markiewicz, and A. Bansil,
%Competing order scenario of two-gap behavior in hole doped cuprates.
{Phys. Rev. B} {\bf 77}, 134516 (2008); T. Das, R. S. Markiewicz, A. Bansil, and A. V. Balatsky,
%Visualizing electron pockets in cuprate superconductors, {\it Preprint available at http://http://arxiv.org/abs/1203.5746}.
Phys. Rev. B {\bf 85}, 224535 (2012).
%
\bibitem{fldCO}T. Wu {\it et al.},
%Magnetic field induced charge-stripe order in the high-temperature superconductor YBa$_2$Cu$_3$O$_y$.
{Nature} {\bf 477}, 191 (2011).
%
\bibitem{Harrison}N. Harrison, and S. E. Sebastian, %Protected nodal electron pocket from multiple-$Q$ ordering in underdoped high temperature superconductors,
{Phys. Rev. Lett.} {\bf 106}, 226402 (2011).
%
\bibitem{Sachdev_SDW}A. Hackl, and S. Sachdev,
%Nernst effect in the electron-doped cuprates.
{Phys. Rev. B} {\bf 79}, 235124 (2009);
%
\bibitem{DDW_Nernst}C. Zhang, S. Tewari, and S. Chakravarty,
%Quasiparticle Nernst effect in the cuprate superconductors from the $d$-density wave theory of the PG phase.
{Phys. Rev. B} {\bf 81}, 104517 (2010).
%
\bibitem{ARPES_borisenko}V. B. Zabolotnyy {\it et al.},
% PG in the chain states of YBCO. {\it Preprint available at http://arxiv.org/abs/1111.4068}.
%,1 A. A. Kordyuk,1 D. Evtushinsky,1 V. N. Strocov,2 L. Patthey,2 T. Schmitt,2 D. Haug,3 C. T. Lin,3 V. Hinkov,4, 3 B. Keimer,3 B. Bchner,1 and S. V. Borisenko1
arXiv:1111.4068.
%
\bibitem{STM_FA}Y. Kohsaka {\it et al.},
% How Cooper pairs vanish approaching the Mott insulator in Bi$_2$Sr$_2$CaCu$_2$O$_{8+\delta}$.
{Nature} {\bf 454}, 1072 (2008).
%Y. Kohsaka, C. Taylor, P. Wahl, A. Schmidt, Jhinhwan Lee, K. Fujita, J. W. Alldredge, Jinho Lee, K. McElroy, H. Eisaki, S. Uchida, D.-H. Lee, J. C. Davis
%
%
\bibitem{Hussey}N. E. Hussey {\it et al.} Phys. Rev. Lett. {\bf 80}, 2909 (1998); N. E. Hussey {\it et al.} Phys. Rev. B {\bf 61}, 6475(R) (2000). 
%
\bibitem{Riggs}S. C. Riggs {\it et al.} Nat. Phys. {\bf 7}, 332 (2011).
%, O. Vafek, J. B. Kemper, J.B. Betts, A. Migliori, W. N. Hardy, Ruixing Liang, D. A. Bonn, G.S. Boebinger
%
\bibitem{atkinson}W.A. Atkinson,
%Disorder and chain superconductivity in YBa$_2$Cu$_3$O$_{7-\delta}$.
{Phys. Rev. B} {\bf 59}, 3377 (1999).
%
\bibitem{footortho}With doubling the tetragonal unit cell along the x-axis, we recover the orthorhombic-II phase of this system, and the computed electronic states in both cases match with each other and also with the first-principle calculations.\cite{LDA} Note that when the periodicity is imposed on the chain state via considering orthorhombic unit cell, the open-orbit chain FS transforms into closed pockets.
%
\bibitem{footTB}The intra-planer and inter-planer dispersion is given by $\xi_{pp} =-2t(\phi_x+\phi_y) -4t^{\prime}\phi_x \phi_y -2t^{\prime\prime}(\phi_{2x}+\phi_{2y})-\mu_p,$ and $\xi_{pp^{\prime}} =-2t_{pp^{\prime}}(\phi_x-\phi_y)\phi_{z/4},~~~\xi_{cp} =-2t_{cp}\phi_{z/4},$, wheras the hopping between plane and chain state is obtained as $\xi_{cc} =-2t_{c}\phi_y-\mu_c$. Here $\phi_{\alpha \nu}$=$\cos{(\alpha k_{\nu})}$ with $\nu$=$x$,$y$,$z$. $\mu_{p,c}$ are the chemical potential for the plane and chain states, respectively, which encode relative onsite energy differences and other crystal effects between the two levels. $t$,$t^{\prime}$, and $t^{\prime\prime}$ are the 1st, 2nd and 3rd nearest neighbor (NN) hopping matrix-elements on the 2D CuO$_2$ plane, $t_c$ is the NN hopping on the 1D chain aligned along $y$-axis, and $t_{pp^{\prime}}~(t_{cp})$ is the plane-plane (chain-plane) tunneling matrix-element along $c$-axis. See Supplementary Material\cite{SM} for details.
%
\bibitem{MookN}B. Fauqu\"e {\it et al.},
%Y. Sidis1, V. Hinkov2, S. PailhÃ¨s1,3, C. T. Lin2, X. Chaud4, and P. Bourges1,*
Phys. Rev. Lett. {\bf 96}, 197001 (2006); H. A. Mook {\it et al.},
% Y. Sidis,2 B. FauquÃ©,2 V. BalÃ©dent,2 and P. Bourges2
Phys. Rev. B {\bf 78}, 020506(R) (2008).
%
\bibitem{GrevenN}Yuan Li {\it et al.},
% V. BalÂ´edent,2 N. BariË‡siÂ´c,3,4 Y. C. Cho,5 Y. Sidis,2 G. Yu,6 X. Zhao,3,7 P. Bourges,2 and M. Greven
Phys. Rev. B {\bf 84}, 224508 (2011).
%
\bibitem{MookmuSR}J. E. Sonier {\it et al.},
% V. Pacradouni,1 S. A. Sabok-Sayr,1 W. N. Hardy,2 D. A. Bonn,2 R. Liang,2 and H. A. Mook
Phys. Rev. Lett. {\bf 103}, 167002 (2009).
%
%
\bibitem{Dahm}T. Dahm {\it et al.}, %  Strength of the spin-fluctuation-mediated pairing interaction in a high-temperature superconductor.
%V. Hinkov, S. V. Borisenko, A. A. Kordyuk, V. B. Zabolotnyy, J. Fink, B. Bchner, D. J. Scalapino, W. Hanke, B. Keimer
{Nat. Phys.} {\bf 5}, 217 (2008).
%
\bibitem{SM}See Supplementary Material for the detailed derivation of Hall and Nernst coefficients and tighi-binding hopping parameters.
%
\bibitem{JMesot}J. Chang  {\it et al.}
%Y. Sassa, S. Guerrero, M. Mansson, M. Shi, S. Pailhes, A. Bendounan, R. Mottl, T. Claesson, O. Tjernberg, L. %Patthey, M. Ido, N. Momono, M. Oda, C. Mudry, J. Mesot
%Electronic structure near the 1/8-anomaly in La-based cuprates.
{New J. Phys.} {\bf 10}, 103016 (2008).
%
\bibitem{ZXshadow} R.H. He {\it et al.},
%X. J. Zhou, M. Hashimoto, T. Yoshida, K. Tanaka, S.-K. Mo, T. Sasagawa, N. Mannella, W. Meevasana, H. Yao, M. %Fujita, T. Adachi, S. Komiya, S. Uchida, Y. Ando, F. Zhou, Z. X. Zhao, A. Fujimori, Y. Koike, K. Yamada, Z. %Hussain, Z.-X. Shen
%Doping dependence of the $(\pi,\pi)$ shadow band in La-based cuprates studied by angle-resolved photoemission spectroscopy.
{New J. Phys.} {\bf 13}, 013031 (2011).
%
\bibitem{RelaxA}J.M. Ziman,
%Electrons and Phonons
{Oxford University Press, Oxford} (1960).
%
\bibitem{foottau} Since the conductivity tensors are proportional to the relaxation time, this particular choice of featureless $\tau$ has no major influence on the overall shape of the coefficients.
%
\bibitem{TdepPG}The values of $R_H$ at $T=0$ and $T_s$ are constrained by the hole doping concentration and density wave transition temperature, respectively, while its $T$-evolution between them can be well reproduced by assuming a competing order origin of pseudogap with the gap function $V(T)=V(0)(1-T/T_s)^r$ with the critical exponent $r=0.5$ as obtained within BCS theory. This $T$-evolution is also used in earlier theoretical studies\cite{Sachdev_SDW} as well as to fit experimental the data [D. D. Prokof'ev, M. P. Volkov, and Yu. A. Boikov, Phys. Solid State, {\bf 45}, 1223 (2003)].
%
\bibitem{Ts}N. Ichikawa {\it et al.},
%Ichikawa, N., {\it et al.}
%Local magnetic order vs superconductivity in a layered cuprate.
{Phys. Rev. Lett.} {\bf 85}, 1738 (2000).
% S. Uchida1, J. M. Tranquada2, T. Niemller3, P. M. Gehring4, S.-H. Lee4,5, and J. R. Schneider3
%
\bibitem{Nernst_prepairing}K. Behnia,
%Behnia, K. The Nernst effect and the boundaries of the Fermi liquid picture.
{. Phys. Condens. Matter} {\bf 21}, 113101 (2009).
%
\bibitem{ambipolar}V. Oganesyan, and I. Ussishkin,
%Nernst effect, quasiparticles, and $d$-density waves in cuprates.
{Phys. Rev. B} {\bf 70}, 054503 (2004).
%
\bibitem{Sondheimer}E.H. Sondheimer,
%Sondheimer, E.H.
%The theory of the galvanomagnetic and thermomagnetic effects in metals.
{Proc. R. Soc. London, Ser. A} {\bf 193}, 484 (1948).
%
\bibitem{Nernst_data_foot}We have chosen the low-field Nernst effect data from Ref.~\onlinecite{Nernst_data} for comparison, since the other experimental data near $H_{c2}$ is subject to acquire contributions from vortex induced fluctuating pairs, which is not included in our theory.
%
\bibitem{transport_nematic}Y. Ando {\it et al.},
%Ando, Y. {\it et al.}
%Electrical resistivity anisotropy from self-organized one dimensionality in high-temperature superconductors.
%K. Segawa, S. Komiya, and A. Lavrov,
{Phys. Rev. Lett.} {\bf 88}, 137005 (2002).
%
\bibitem{HinkovScience}V. Hinkov {\it et al.},
% D. Haug,1 B. FauquÃƒÆ’Ã‚Â©,2 P. Bourges,2 Y. Sidis,2 A. Ivanov,3 C. Bernhard,4 C. T. Lin,1 B. Keimer1
%Electronic Liquid Crystal State in the High-Temperature Superconductor YBa$_2$Cu$_3$O$_{6.45}$.
{Science} {\bf 319}, 597 (2008).
%
\bibitem{EAKNature} M. J. Lawler {\it et al.},
%Lawler, M. J. {\it et al.} Intra-unit-cell electronic nematicity of the high-$T_c$ copper-oxide PG states.
% K. Fujita2,3,4*, Jhinhwan Lee2,3,5, A. R. Schmidt2,3, Y. Kohsaka6, Chung Koo Kim2,3, H. Eisaki7, S. Uchida4, J. C. Davis2,3,8, J. P. Sethna2 & Eun-Ah Kim2
{Nature} {\bf 466}, 347 (2010).
%
\bibitem{refael}T. Pereg-Barnea {\it et al.},
%Pereg-Barnea, T. {\it et al.}
%Quantum oscillations from Fermi arcs.
{Nat. Phys.} {\bf 6}, 44 (2009).
%Weber, G. Refael1 and M. Franz2
%
\bibitem{open_orbit_QO}J.M. Tranquada {\it et al.},
%Tranquada, J.M., {\it et al.} Reconsidering the interpretation of quantum oscillation experiments on underdoped YBa$_2$Cu$3$O$_{6+x}$.
{Phys. Rev. B} {\bf 81}, 060506(R) (2010).
%, D. N. Basov, A. J. LaForge, A. A. Schafgans
%
%
\bibitem{DasINS}T. Das, %In-plane anisotropy in spin-excitation spectra originating from chain states in YBa$_2$Cu$_3$O$_{6+y}$.
{Phys. Rev. B} {\bf 85}, 144510 (2012).
%
\bibitem{QOY124}P. M. C. Rourke {\it et al.},
%A. F. Bangura, C. Proust (2 and 3), J. Levallois (2), N. Doiron-Leyraud (4), D. LeBoeuf (4), L. Taillefer (3 and 4), S. Adachi (5), M. L. Sutherland (6), N. E. Hussey
Phys. Rev. B {\bf 82}, 020514(R) (2010).
%
\bibitem{Hudson}W. D. Wise {\it et al.}
% Kamalesh Chatterjee, M. C. Boyer, Takeshi Kondo, T. Takeuchi, H. Ikuta, Zhijun Xu, Jinsheng Wen, G. D. Gu, Yayu Wang, E. W. Hudson
Nat. Phys. {\bf 5} 213-216 (2009).
%
\bibitem{RIXS}G. Ghiringhelli, 
% M. Le Tacon, M. Minola, S. Blanco-Canosa, C. Mazzoli, N. B. Brookes, G. M. De Luca, A. Frano,
%    * D. G. Hawthorn, F. He, T. Loew,* M. Moretti Sala,* D. C. Peets,    * M. Salluzzo,     * E. Schierle,    * R. Sutarto,     * G. A. Sawatzky,     * E. Weschke,     * B. Keimer,     * and L. Braicovich
Science, Published online 12 July 2012 [DOI:10.1126/science.1223532]; arXiv:1207.0915.
%
\end{thebibliography}

\begin{thebibliography}{10}
\bibitem{markietb}R. S. Markiewicz {\it et al.} 
%One-band tight-binding model parametrization of the high-$T_c$ cuprates including the effect of kz dispersion. 
{Phys. Rev. B} {\bf 72}, 054519 (2007).
%, S. Sahrakorpi, M. Lindroos, Hsin Lin, and A. Bansil
%
\bibitem{atkinson}W. A. Atkinson, 
%Disorder and chain superconductivity in YBa$_2$Cu$_3$O$_{7-\delta}$. 
{Phys. Rev. B} {\bf 59}, 3377 (1999).
%
\bibitem{DasSDW}T. Das, R. S. Markiewicz, and A. Bansil,
% Competing order scenario of two-gap behavior in hole doped cuprates. 
{Phys. Rev. B} {\bf 77}, 134516 (2008).
%
\bibitem{DDW}S. Chakravarty, and H.Y. Kee. % Fermi pockets and quantum oscillations of the Hall coefficient in high-temperature superconductors. 
{Proc. Nat. Acad. Sci. USA} {\bf 105}, 8835 (2008).
%
\bibitem{Sachdev_stripe}A. Hackl, M. Vojta, and S. Sachdev, 
%Quasiparticle Nernst effect in stripe-ordered cuprates. 
{Phys. Rev. B} {\bf 81}, 045102 (2010).
%
\bibitem{Ts}N. Ichikawa {\it et al.},
% Local magnetic order vs superconductivity in a layered cuprate. 
{Phys. Rev. Lett.} {\bf 85}, 1738 (2000).
% S. Uchida1, J. M. Tranquada2, T. Niemller3, P. M. Gehring4, S.-H. Lee4,5, and J. R. Schneider3
%
\bibitem{Nernst_prepairing} K. Behnia,
%The Nernst effect and the boundaries of the Fermi liquid picture. 
{J. Phys. Condens. Matter} {\bf 21}, 113101 (2009).
%
\bibitem{ambipolar}V.  Oganesyan, and I. Ussishkin, 
%Nernst effect, quasiparticles, and $d$-density waves in cuprates. 
{Phys. Rev. B} {\bf 70}, 054503 (2004).
%
\bibitem{Sachdev_SDW}A. Hackl, and S. Sachdev, 
%Nernst effect in the electron-doped cuprates. 
{Phys. Rev. B} {\bf 79}, 235124 (2009).
%
\bibitem{DDW_Nernst}C. Zhang, S. Tewari, and S. Chakravarty, 
%Quasiparticle Nernst effect in the cuprate superconductors from the $d$-density wave theory of the pseudogap phase. 
{Phys. Rev. B} {\bf 81}, 104517 (2010).
%
\bibitem{Nernst_QSH}D.I. Pikulin, C.Y. Hou, and C.W.J. Beenakker, 
%Nernst effect beyond the relaxation-time approximation. 
{hys.Rev. B} {\em 84}, 035133 (2011).
%
\bibitem{RelaxA}J. M. Ziman, 
%Electrons and Phonons 
{Oxford University Press, Oxford} (1960).
%
\bibitem{HallEP}D. LeBoeuf {\it et al.},
% Electron pockets in the Fermi surface of hole-doped high-$T_c$ superconductors. 
{Nature} {\bf 450}, 533-536 (2007).
%David LeBoeuf1, Nicolas Doiron-Leyraud1, Julien Levallois2, R. Daou1, J.-B. Bonnemaison1, N. E. Hussey3, L. Balicas4, B. J. Ramshaw5, Ruixing Liang5,6, D. A. Bonn5,6, W. N. Hardy5,6, S. Adachi7, Cyril Proust2 & Louis Taillefer
%
\bibitem{Nernst_data}J. Chang {\it et al.} 
% Nernst effect in the cuprate superconductor YBCO: Broken rotational and translational symmetries. 
{Phys. Rev. B} {\bf 84}, 014507 (2011).
%J. Chang, Nicolas Doiron-Leyraud, Francis Lalibert, R. Daou, David LeBoeuf, B. J. Ramshaw, Ruixing Liang, D. A. Bonn, W. N. Hardy, Cyril Proust, I. Sheikin, K. Behnia, Louis Taillefer
%
\bibitem{Vojta}A. Hackl, and M. Vojta, Phys. Rev. B {\bf 80}, 220514(R) (2009).
%
\end{thebibliography}
\end{document}